# The Role of Translated Information Quality in a Global e-Retailing Context

**Wanxian Zeng**
Research School of Management
The Australian National University
Canberra, Australia
Email: u5516808@anu.edu.au

**Alex Richardson**
Research School of Management
The Australian National University
Canberra, Australia
Email: alex.richardson@anu.edu.au

## Abstract

Global e-retailing continues to soar in popularity, but scant attention is being paid to the role of translation. This paper proposes a study investigating whether improving translated information quality of product descriptions increases the customers' information satisfaction, while reducing the perceived product risk, which in turn improves their intention to use an online shopping website. To manipulate translation quality, two translation methods are used: machine and crowdsourced. The chosen translation written language pair is from English to Simplified Chinese, as these are the official languages of the two largest economies (U.S.A and China respectively) that also have large e-tailing markets. A model based on an integration of two theories, DeLone & McLean's Information Systems Success Model, and Perceived Risk Theory, has been developed for testing the impact of translated information quality. The moderating effect of translation method and e-retailer's brand influencing customers' tolerance of imperfect translation is also considered.

## 1   Introduction

Over the past decade, e-retailing has evolved from a convenient novelty to the platform for global expansion and the largest driver of retail sales growth (Devitt et al. 2013). According to eMarketer (2014), global e-retailing sales reached US$1.25 trillion in 2013, accounting for 5.9% of total retailing sales, and sales grew at a rate of 20% per annum (on average) during 2012 and 2013, led by China with 86% growth. Additionally, the sales of cross-border e-retailing reached US$300 billion globally in 2012 (CBEC 2015) and are expected to soar to more than 30% of global online trade overall by 2020 (IMRG 2015). As companies expand into global markets, their customer base becomes more diverse and spans multiple countries in which people communicate in different languages (Levitt and Limburg 2011). Consequently, language diversity can become a barrier when e-retailers enter a foreign market, necessitating translation and the associated costs. According to DePalma (2006), the possibility of a successful online transaction is four times higher when the website content is presented in a customer's native language. Furthermore, even among the multilingual who also understands the original source language, customers' preferences for local/native lanuage content is evident across countries (Gibbs et al. 2003). On the other hand, mistranslation will cause business loss. In 2009, HSBC Bank launched a US$10 million rebranding campaign to correct the mistake arising from its catchphrase "Assume Nothing" being mistranslated in many countries to "Do Nothing" (Vincent 2009).

While the demand for translation is increasing, services provided by professional service firms are expensive and time-consuming. ProZ.com, a website that lists the rates for translation services from its community of translation companies and freelance translators, reports that the average rate (based on the 4,781 samples) of translation from English to Chinese is US$0.11 per word (ProZ 2015). While contracts and agreements warrant the need for professional translation quality that outweigh the costs, many written pieces cannot justify the costs of professionalism where the benefits are minimal. To manage the language barrier at a lower cost, two potential approaches are utilised: 1) machine

translation (MT) which is the use of computerised systems to translate between languages (e.g. Google Translate); and 2) crowdsourced translation which obtains translations by gathering contributions from an extensive range of bilinguals, particularly from online communities (e.g. Gengo). The quality of MT and crowdsourced translation is improving quickly, and is changing the face of the translation industry by providing any business the opportunity to outsource translation tasks at an affordable price point. For example, Google Translate launched a project in July 2014 to improve its translation service by crowdsourcing translation knowledge directly from its users by having them provide corrections for mistranslations. This crowdsourced translation will influence Google's algorithms, and allow the translation system to learn languages better (Lardinois 2014). In addition, online translation platforms, such as Gengo or Get Localization, amplify the power of crowdsourcing by using the Internet to facilitate access to cheaper and quicker translation services. These two translation methods are selected in the manipulation of translation quality in this research as they are potential alternatives to professional translation services.

This study focuses on written language translation from UK/US English to Simplified Chinese because China is the largest global market for e-retailing according to the number of online customers and also the revenue (UNCTAD 2015). In 2014, the number of Chinese Internet users reached 649 million (McKirdy 2015), with 361 million of them being online shoppers – an increase of 19.7% compared with the previous year (Knowler 2015). Furthermore, 2013 saw online retail sales in China totalling US$296.57 billion, an increase of 41.2% over 2012 — triple the growth rate of overall retail sales (Tong 2014). In terms of cross-border e-retailing, 18 million Chinese customers spent US$35 billion on cross-border online shopping in 2013 (Chen 2015). This research is timely as China recently allowed full foreign ownership of e-commerce companies. Previously, foreign investors required a Chinese partner to enter the Chinese online shopping market, and their partnership ownership stake needed to be below 55% (Liu 2015). This change indicates that the Chinese government is giving more support to online cross-border trade, which is expected to increase the cross-border e-retailing sales in the coming years. It provides an attractive opportunity for English speaking companies to sell their goods to the Chinese market, provided they can effectively communicate in order to conduct business.

The aim of this study is to examine the role of translated information quality in the success of a global e-retailing system. The focus is on the information quality, where the potential customer only has the translated version of the information presented to them. Moreover, it investigates whether the quality of information translated by the more cost effective methods of MT and crowdsourced translation are useful and acceptable to online customers in an e-retailing context. Furthermore, the research investigates whether brand awareness can moderate the relation between translated information quality and perceived product risk. We propose the following research question:

*What is the relationship between the translated information quality and potential customers' intention to use a particular shopping website in the global e-retailing context?*

This research-in-progress paper proceeds as follows. First, the relevant literature reviewed and how it informs the hypothesis development is explained. Then the proposed research method is discussed, followed by a brief overview of future plans and potential contributions.

## 2    Literature Review

### 2.1    Information Systems (IS) Success Model & Information Quality

Prior literature does not provide a model for using translated information quality as an independent variable, but we present an argument that the well-established, rigorous and validated Information Systems (IS) Success Model is a fitting basis for evaluating the impact of translated information quality. In 1992, after an inclusive review of assessments of IS success and inspired by communication theory (Shannon and Weaver 1949) and information influence theory (Mason 1978), DeLone & McLean proposed a model of interrelations between six IS success variable categories: system quality, information quality, IS use, user satisfaction, individual impact and organization impact. In 2003, DeLone and McLean presented an Updated IS Success Model in response to suggestions they received after the publication of the IS Success Model and also due to the dramatic advance in IS practice, particularly the appearance of e-commerce. Unlike users of traditional ISs who "choose" to use ISs as an employment requirement, customers' use of online shopping websites in e-commerce context is often voluntary. Therefore, "intention to use" is considered as an important measure of IS success in the updated model. In addition, "individual impact" and "organizational impact" have been replaced with a more general construct "net benefits", as ISs have evolved to affect a wider range of people instead of only immediate users (Delone and McLean 2003). Put succinctly, the model contributes to the knowledge of IS success because it presents a structure for classifying different IS success measures in extant literature. Additionally, it suggests a model of temporal and causal relationships between those groups of measures (Wang 2008).

The Updated IS Success Model indicates that information quality, system quality and service quality jointly influence user satisfaction and their intention to use a system. As this research involves information quality, we focus on the part of the IS Success Model that depicts the constructs related to information quality, while system quality and service quality will be controlled for. Our research model does not include net benefits because the purpose is to investigate whether translated information quality will influence customer intention to use the website in the first place, rather than examine how much money, time or effort has been saved by the customer when they use an online shopping website.

In an online purchasing environment, information quality refers to "the usefulness of the available information about attributes of a product", in terms of facilitating an online customer to evaluate that product (Gao et al. 2012). In the IS Success Model, information quality has been found to be one of vital predictors of system users' satisfaction. The information quality and the degree to which it meets the requirements and expectations of customers would affect the success of e-retailers and determine whether a customer will use a particular e-retailer or turn to a competitor (Molla and Licker 2001).

Selling to foreign (language) customers necessitates that information goes through a translation process. Therefore, we propose that translation quality is a part of information quality, as an aspect of understandability and interpretability. Few studies have addressed the influence of translated information quality on foreign customers' satisfaction, but the importance of language and translation in global e-retailing has been noticed. As previously mentioned, Gibbs et al. (2003) argues that language is identified as an inhibitor of global e-retailing for non-English-speaking customers, particularly in Asia where most of the older generation are not familiar with the spoken and written English language. Low English proficiency restricts the growth of cross-border online shopping in China, although the Chinese show a strong interest in foreign products as they are the "top spending tourists" worldwide (Borderfree 2013). Furthermore, even among the multilingual who speak English, there remains a preference for digesting content in their local language (Gibbs et al. 2003). They state that the better the content online is, particularly in the local language, the more it meets the requirements and expectations of customers. In contrast, if the website does not present in the language that customers understand, it will affect their comprehension and understanding. Consequently, customers cannot easily form trust in a website that is in an unknown language.

For assessment of the translated information quality within this research, we draw upon the work of Williams (2004) who summarized two categories of assessment methods: quantitative-centred systems and argumentation-centred systems. The quantitative-centred method is characterized by error counting. Schiaffino and Zearo (2005) suggest that when measuring translation quality, evaluators actually measure the occurrence of different types of errors in the target text (i.e. translations). There are three types of errors:

1) Errors of meaning, i.e. the meaning of source text is different from that of the target text;

2) Errors of form, such as an error of grammar, spelling or other formal error. Those errors do not change the meaning of the translation, but make it harder to fully and correctly understand the original text;

3) Errors of compliance, which happen when the translation fails to follow the style guide, instructions received, preferred terminology or other customer-specified requirements.

Errors are categorized according to different levels, which influences the error weight. Next, detected errors are summed and a numerical score is used to estimate the overall quality level of translation. On the other hand, based on the functionalist perspective (the function of translation is to communicate), Williams (2004) claims that readers' demands and expectations set by a translation buyer must be taken into consideration. It is the reader and the translation buyer's requirement that determines the acceptability of the translations. Williams believes that translation evaluation should not be based on quantitative models of error assessment. Instead, the priority is to obtain a translation that is acceptable for a translation buyer, no matter whether it is "completely flawless or contains any minor mistakes". Nerudová (2012) confirmed that the objectivity measurement (i.e. quantitative-centred approach) of translation quality is not commonly applicable in the business context, where translation buyers have diverse requirements and they have highly subjective assessments of translation and its acceptability.

The distinction between methods is important to acknowledge as it means the way translation experts assess translations is different from how customers do. Therefore, there is possibility that a low quality translation assessed by an expert might still be acceptable for a customer. This means professional translation is not always necessary. Translation service providers like Gengo use the quantitative-centred approach to decide whether a person is qualified to work for them. In addition, after a qualified translator completes a job for customers, Gengo uses software and senior editors to score translators based on error count, error type, error severity and word count (the score is private to translators and Gengo only). Additionally, Gengo asks customers to score its translators and this score is made public to their customers. However, the method customers use is arguementation-centred and concerned with whether the translation is acceptable to customers.

## 2.2 Perceived Risk Theory

Although the IS Success Model shows the impact of information quality on user satisfaction and intention to use, it does not model whether it can influence customers' beliefs in the product itself. Research in the marketing field indicates other factors are closely related to customers' perception towards the product and the e-retailer, e.g. perceived risk. Attention to customers' perceived risk in marketing research was aroused after the work of Bauer (1960), who first suggested that customer behaviour could be regarded as "an instance of risk taking". He differentiates the objective risk and the perceived risk. Researchers should focus on the perceived one since customers will only react to the risks they notice (Glover and Benbasat 2010). Thereafter, perceived risk theory has been widely adopted by scholars in marketing to interpret customer behaviour. Cunningham (1967), cited in Mitchell (1992), suggests that risk comprises two dimensions: uncertainty and consequences. Perceived risk is generated if a person involves in situations where the outcomes are uncertain, and the person is worried about the consequences of a poor or wrong decision. In the context of retailing, the consequence results from a mismatch between actual purchase performance and expected purchase performance (Huang et al. 2004). Glover and Benbasat (2010) suggest IS researchers still need to focus efforts to learn customer' perceived risks, even though the marketing research of risk is well-established, as some risks are magnified in an e-commerce context and highly influential in the customers' decision making. The main reason is that online transactions involve more uncertainty than purchases in physical stores due to factors such as no physical observance of products and no access to a human "face" for the business.

Previous research addresses many different types of risks. Jacoby and Kaplan (1972) recognize five types of perceived risk (i.e. performance risk, financial risk, psychological risk, social risk and physical risk) and their research shows that those five types of risks can explain 74% of the variance in the overall perceived risk measures. In the context of e-retailing, Glover and Benbasat (2010) summarise previous research and find that risks fall into three stages of an e-retailing transaction: 1) phenomenon as the "source" of the risk; 2) an "event" from the source that exposes customers to harm; 3) "one or more types of harm" resulted from an event. Using this idea to analyse the perceived product risk (Chen et al. 2015; Park et al. 2004), it can be concluded that the source of risk is the purchased product; event that may generate harm is performance risk; and harm may occur from the source is financial risk, psychological risk and time risk. In this research, the perceived product risk shares the similar definition as that in Chen et al. (2015) and Park et al. (2004). The perceived product risk results from the product, and its functional risk due to: 1) low quality of products; 2) customers' requirements and the actual product not matching; and 3) low value of the product. In addition to these performance, financial, and psychological risk types, this research also covers the social risk as the experiment task involves buying products for others, and physical risk as the product used in experiment was related to health and well-being.

Because of the pre-purchase intangibility of products, online customers are typically not sure that the product really meets their needs. Furthermore, if the product information that customers can obtain from the shopping website is insufficient, their doubt about the product quality increases, which then decreases their intention to interact further with that e-retailer (Chen et al. 2015). In general, people are prone to avoiding mistakes rather than maximizing utility when engaging in uncertainty (Liao et al. 2010; Mitchell et al. 1999). Glover and Benbasat (2010) confirm "failure to gain product benefit risk", meaning unable to meet expectation, changes customers' attitude and their intention to buy or use an e-retailing store. Purchasing from a foreign brand will further complicate customer perceptions as what constitutes standard product information varies between countries. E-retailers need to be wary of the information requirements of all foreign customers (Chen et al. 2015). However, based on the research of Moshrefjavadi et al. (2012) and Sinha (2010), the impact of perceived product risk on the intention to shop from a specific website is not significant. Therefore, whether e-retailers should put effort into lowering customers' perceived product risk to stimulate their interest in using the online store needs further investigation.

To sum up, our approach synthesizes the Updated IS Success Model path related to information quality, and according to Perceived Risk Theory, captures how information influences customers' perceived risk towards the product quality. Table 1 summarises the function and overlap of the IS Success Model and Perceived Risk Theory in this research.

|  | IS Success Model | Perceived Risk Theory |
| --- | --- | --- |
| Root disciplines | Information system | Psychology; Marketing |
| Purpose | To provide an inclusive framework of IS success by identifying, defining the most critical dimensions of IS success assessment, and explaining the relationships among those dimensions. | To construct a comprehensive frame of risk taking in customer buying behaviour by specifying the principal risk types involved and how they influence every stage of customer decision-making process. |
| Relevant core constructs | Information quality; user satisfaction; intention to use. | Perceived product risk. |
| Role in research model | To examine the factors that influence how foreign customers form their intention to build relationship with a particular e-retailer by assessing the information provided by online stores. | To extend the information quality part of IS Success Model by considering the effect of perceived risk on the intention to use a particular website. |

*Table 1. Summary of Research Models informing this Study*

## 2.3 Research Model and Hypotheses Development

From the literature discussed in the previous sections, a research model has been developed for examining customers' online shopping behaviour in a global e-retailing context where translated information quality needs to be considered. The model adds a new construct "Perceived Product Risk", which has been considered as an important factor to evaluate the intention to use (Chen et al. 2015; Wafa 2009). For the key moderating factor affecting the potential customers' perceived product risk, we examine "Brand awareness". With the aim of an incremental improvement in understanding, all constructs and model paths are derived and informed respectively from an extensive range of previous e-commerce and marketing research. The resulting research model diagram is presented in Figure 1, which is then followed by the hypotheses development discussion.

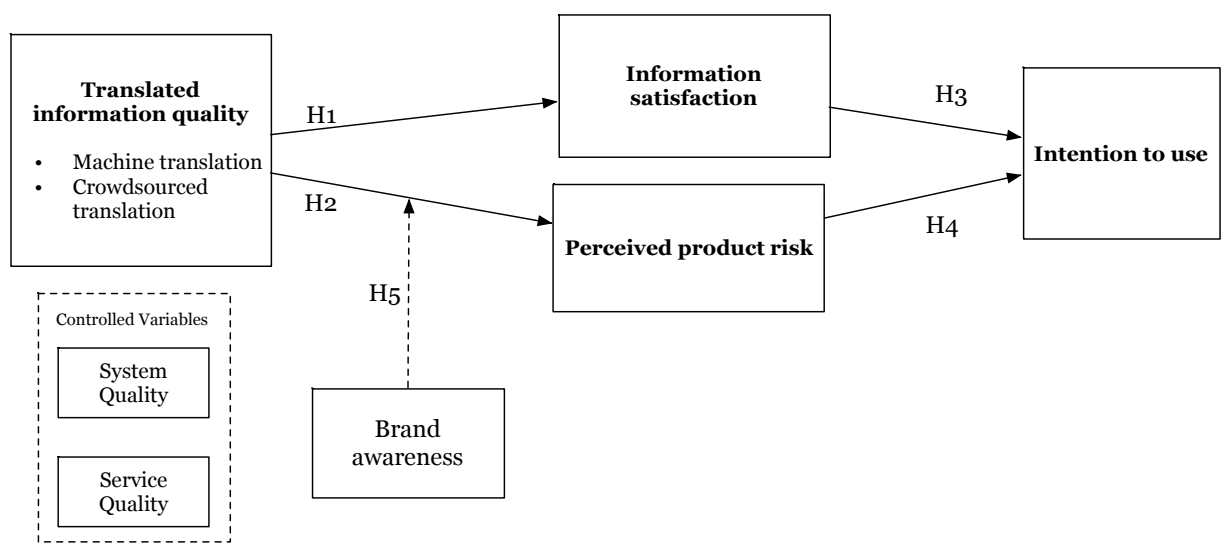

*Figure 1: Research Model*

### 2.3.1 Translated Information Quality and its Impact

The hypothesized relationship between the translated information quality of product information and information satisfaction is based on the theoretical and empirical work reported by DeLone and McLean (2003; 2004). The Updated IS Success Model shows information quality is positively related to user satisfaction. Wixom and Todd (2005) separate IS Success Model user satisfaction into information satisfaction and system satisfaction, to which Xu et al. (2013) adds service satisfaction. As the purpose of this research is to address the impact of information quality, the satisfaction is derived from the product information provided to the user (i.e. the e-tailer customer). Both Wixom and Todd (2005) and Xu et al. (2013) suggests that information quality positively impacts customer's information satisfaction. Corresponding with the IS Success Model, several studies in the field of marketing also suggest that information quality is the antecedent of customer satisfaction (Chae et al. 2002; Rodgers et al. 2005; Spreng et al. 1996). Through acquiring information related to the product, customers can establish confidence in realising the expected benefits of their purchases. It is surmised that the higher the translated information quality (of product information) is, the higher customer information satisfaction will be.

> *H1 - Translated information quality is positively related to potential customers' information satisfaction.*

In an e-retailing context, customers face a high degree of perceived product risk due to not being able to examine the product before purchase. Additionally, unknown foreign brands from foreign markets are believed to increase the level of risk perceived (Riegelsberger et al. 2003). When customers shop online, they cannot easily gain comprehensive knowledge about the product quality, so that the purchase decision is largely based on available information provided by the e-retailer. Acquiring better quality information reduces uncertainties regarding the electronic transaction and hence decreases the related perceived risk, because of the worth of the product quality conveyed by the information (Nicolaou et al. 2013). For foreign customers who rely upon translated information due to language disparity, different translation methods will influence information quality in terms of understandability and interpretability.

> *H2 - Translated information quality is negatively related to potential customers' perceived product risk.*

### 2.3.2 Direct Determinants of Intention to Use

Information Satisfaction is the extent to which the user agrees the retrieved information is useful. The Updated IS Success Model shows user satisfaction influencing the intention to use an information system. Waal et al. (2012) verify this model and find the satisfaction has a strongly positive impact on the intention to use. Wixom and Todd (2005) show information satisfaction is part of user satisfaction, and Park and Kim (2003) state that information satisfaction plays a key role in forming site commitment. Devaraj et al. (2002) measures customer (user) satisfaction in the e-retailing context and supports empirically that satisfaction is a key antecedent of customers' channel preference. Furthermore, the body of literature supports the stance that satisfied customers have a greater intention to use company's products, a greater re-purchase intention, more positive word-of-mouth, and a lower propensity to look for alternative product suppliers (Wang 2008).

> *H3 - Potential customers' information satisfaction is positively related to their intention to use a particular e-retailing website.*

Perceived risk theory is well-suited in explaining "how a perceived risk directly influences intention" (Mitchell et al. 1999). The perceived product risk is a particularly important dimension in a customer's online purchasing decision (Luo et al. 2012) and has been considered a major barrier to online transactions (Gefen et al. 2008). The uncertainty about the product quality and the consequence of buying unwanted product will decrease potential customers' intention to buy from that online stores as "people are prone to avoid mistakes rather than maximize utility when engage in uncertainty" (Liao et al. 2010; Sheth and Venkatesan 1968). It has been found that high levels of perceived product risk are related to low levels of positive feelings during the consumption experience and decreases the customers' evaluation of overall performance (Chaudhuri 1998; Kim et al. 2008; Park et al. 2005). When customers perceive high risk, they need to exert more effort to confirm or eliminate their suspicions, which lowers their site commitment and increases the probability of leaving the website without committing to a purchase (Belanche et al. 2012; Park and Kim 2008).

> *H4 - Potential customer' perceived product risk is negatively related to their intention to use a particular e-retailing website.*

### 2.3.3  The Moderating Effect of e-Retailer Brand Awareness

Perceived product risk is expected to lead to risk handling behaviours such as "seeking information, wider search, increased use of word of mouth information sources and buying from well-known, major brands" (Clarke 2007). Due to the proliferation of e-tailing websites, customers can choose from an increasingly greater number of options. In addition to the manufacturer brand of a product (which we control for), the (e-) retailers' brand will also influence the customer's decision of which online shopping website to buy from when the same (or similar) product is sold by various e-retailers. When shopping online, customers look for signals indicating whether the e-retailer is trustworthy. McKnight et al. (2002) proposes that customer knowledge about unknown retailers is more heavily based on first impressions, and their trust of the unknown brand is based on whether the website looks "normal" (Bahmanziari et al. 2009). On the other hand, customers prefer to buy from well-known e-retailers, especially when the manufacturer brand of a product is less well known (Chu et al. 2005). Customers believe that famous e-retailer brands bond their reputation with that of the manufacturer when they sell a product. If those retailers are found to sell low quality products, their reputation will be damaged even though they are not the manufacturer of the low quality product. As a result, in the context of e-retailing, brand awareness of an e-retailer may mitigate customers' perceived risk towards shopping online (Kwon and Lennon 2009), and lessen the negative impact of perceived website quality (Kim and Jones 2009), therefore, making them more confident in buying from that e-retailer' online store (Hongyoun Hahn and Kim 2009) .

> *H5 - The effect of translated information quality on perceived product risk will be reduced when brand awareness of the e-retailer is high (well-known brand) and be higher when brand awareness is low (unknown brand).*

## 3  Proposed Research Method

An experiment will be conducted using a 2x2 within-subjects factorial design, in which the same group of participants serve in all treatment conditions. One of the greatest advantages of using an experimental approach is that it provides a high degree of control and the within-subjects design can help limiting errors arising from individual differences and requires fewer participants and resources. Participants will view fabricated e-retailer webpages (real brand but standardised appearance) containing product information in written Chinese. A vignette will be used to set the context and role of participants as shoppers. For the experiment, brand awareness and translation quality will be manipulated, which results in the 4 treatment conditions:

1. Well Known x Machine Translation
2. Well Known x Crowdsourced translation
3. Unknown x Machine Translation
4. Unknown x Crowdsourced translation

### 3.1  Participants

University students will be a primary source of participants as they are representative of active users of the Internet and online shopping websites, and form a major part of the local community who have a mainland Chinese education and tertiary level Chinese lanuage proficiency. As the age of the majority (63.4%) of online customers in China range from 18 to 30 years old (HKTDC 2014), and therefore the typical university student fits the age range well. However, due to the potential for a narrow characteristics of participants (postgraduate Commerce students with a high education, understanding of more than one language, and low income), it would be hard to generalize to individuals who do not have those characteristics. Therefore, we plan to involve a wider range of participants, i.e. members of the general public and students majoring in different disciplines (including undergraduates). Subjects will receive $10 for their participation to compensate them for the travel expenses and the 30 minutes (approximately) the experiment session will take.

### 3.2  Materials

A well-designed experiment can manipulate the independent variable and control for confounding effects or alternative explanations and test the causal relations (Creswell 2013). To the greatest extent possible, the questionnaire instrument used in the experiment will be based on extant pre-validated items from literature, as intention to use, user (information) satisfaction and perceived risk are all well-established constructs (refer to Table 2 for source).

As far as we know, there are no extant instruments for measuring translated information quality. What customers need to derive from an online shopping website is not the "information" itself, but rather influential factors represented by the information quality (Ghasemaghaei and Hassanein 2015). Hence, the study utilizes the "Representational IQ" dimensions from Lee et al. (2002) as those measures of information quality are a continuation of Wang and Strong (1996). The representational IQ dimensions cover both the format and meaning of data. Specifically, customers conclude whether the information quality is "good" according to the format (concise and consistent representation) and the meaning of data (interpretability and ease of understanding) (Wang and Strong 1996). As translation does not need to affect the format of data (post-processing controls for this), we focus on measuring "understandability" and "interpretability" of the product information, and for measuring the overall information quality of an online store.

The following Table 2 shows the variables and the indicators involved in this study.

| Variable name | Variable type | Variable description | Indicators |
| --- | --- | --- | --- |
| Translated information quality | Independent variable [Reflective] | The quality of product information provide by e-retailers in facilitating product and e-retailer evaluation, which is presented in the translated form. | ·Understandability<br>·Interpretability<br>    (Lee et al. 2002)<br>·Overall information quality<br>    (Xu et al. 2013) |
| Intention to use | Dependent variable [Reflective] | The willingness of participants to use an online store of a specific retailer. | ·Intention to visit the e-retailing website<br>·Intention to use the e-retailing website for information retrieval<br>·Purchase intention<br>    (Delone and McLean 2003)<br>·Intention to continue using the shopping website<br>    (Chen and Cheng 2009) |
| Brand awareness | Moderating variable [Dichotomous] | Whether customers can recognize the brand within various contexts or situations. | ·Aware of the brand<br>·Ability of recalling<br>·Ability of recognizing<br>    (Jinfeng and Zhilong 2009) |
| Information satisfaction | Mediating variable [Reflective] | Customers' satisfaction with information provided by an e-retailer. In this study, it comes from the available information in the product description. | ·Overall information satisfaction<br>·Information satisfaction in terms of meeting expectation<br>·Satisfaction about the website comparing with its counterparts.<br>    (Park and Kim 2003) |
| Perceived product risk | Mediating variable [Reflective] | The uncertainty that customers have about whether their expectations will be achieved due to lack of observation of the product in person before purchasing. | ·Product quality confidence<br>·Lack of feeling of product<br>·Actual value of product<br>    (Chen et al. 2015) |

*Table 2. Variable Descriptions*

While we are directly manipulating the level of brand awareness to be a dichotomous moderating variable (effectively to the two extremes of "unknown" and "well known"), we still incorporate items for brand awareness from Homburg et al. (2010), based on Aaker (1996), to validate the manipulation is effective.

The majority of items for each construct will use Likert seven-point agreement scales for determining the level of agreement. Representative information quality is a 0-10 scale from "Not at all" to "Completely". Apart from the manipulated and measured variables, we will also control for the effects of extraneous variables which could confound our findings, such as system quality, service quality, website appearance, product information, selection and price, and participants' language proficiency.

The experiment design, procedure and questionnaire instruments will be pre-tested to ensure the validity of experiment conduct and data collected. Careful attention will be placed on pre-testing the effectiveness of the translated product information and choice of e-retailer brands. In terms of the translation assessment, the perception of translation quality from translation experts and from the average people might be different (Nerudová, 2012). Translation experts will be consulted for providing an objective evaluation of translation quality resulting from the two different translation methods, by using the quantitative-centred approach to assess each translation. These evaluations will be cross-checked with the data collected from the pre-test, where participants using the argumentation-centred approach (e.g. by checking the usefullness of the translated information), to determine whether the test participants' evaluations correspond with the experts' opinion. Real company names from pre-rest shortlisting will be used for well-known and unknown brands, with the vignette controlling quality perceptions. This manipulation of brand awareness requires testing to ensure the participants are internalising the brand and adjust behaviour in a suitable manner.

### 3.3 Procedure

The experiment will be conducted at a Group of Eight university in a computer lab (or large office) where the computer configurations are identical. Each session is expected to involve five participants simultaneously to expedite data collection but still allow for a high level of control. After reading the information sheet and consent form, the participants will use experiment software to complete a short demographic questionnaire requiring them to report their gender, age, level of online shopping experience, and average online and "off-line" monthly shopping spends. After providing this information, the software will then start the online shopping scenario part of the experiment. No purchases of actual goods will be made.

The task and product information content presented to each participant will be identical, but the order will differ. The vignette will be given in written form and lead participants through a hypothetical task of needing to buy something for their friend to fulfil a certain set of requirements. Page layout, colour and text format will be controlled by using identical webpage designs but having the e-retailer brand prominent at the top to maintain a clear visual distinction between brands. In order to control the product category but also achieve a level of immersion for validity purposes, participants will receive information in each treatment about more than one product type (pre-testing to confirm exact number), all of which have some aspect of service, experience or credence goods, as well as a range of prices (inexpensive to expensive). Nelson (1970) classifies products into search goods and experience goods based on whether, and when, it is possible to ascertain product quality by obtaining product quality information. Search goods are those products characterized by attributes for which product quality is easy to objectively evaluate before making transactions. This category consists of products that are easier to evaluation without examining in person, such as digital cameras and computers, while experience goods have to be assessed or tried personally, such as clothes, shoes and beauty products (Kiang et al. 2011; Suwelack et al. 2011). In addition, Darby and Karni (1973) emphasise the need to add credence goods as the third category, whose quality are "expensive to judge even after purchase" (e.g. vitamin supplements). By providing a range of products that cover all three categories, it will be more realistic for participants to evaluate an e-retailer website than if there was only one product and one category. For each product, we will measure the understandability and interpretability of product information. For each treatment, we administer questions measuring the constructs in our research model. After undergoing all treatments, participants will complete a post-experiment debriefing questionnaire where they may volunteer information about their experience and issues they may have faced.

Because the experiment is a within-subject design, there could be the "carryover effects". In general, this means the participation in one condition may affect performance in other conditions, thus creating a confounding extraneous variable that varies with the independent variable (Hall 1998). To address this issue, a counterbalancing design will be applied. Experiments conducted with a counterbalanced measures design are one of the best ways to avoid the pitfalls of standard repeated measures designs, where the subjects are exposed to all of the treatments (Shuttleworth 2009). In this study, the four possible conditions require 24 orders of treatments (4x3x2x1), which means the number of participants in total will be a minimum and multiple of 24 (if ensuring equal assignment to each treatment order). Participants will be given different orders of treatment and will not be allowed to go back to change their responses to a previous treatment. Furthermore, to reduce the effect of memorizing content knowledge gained from previous treatments, we will prepare translated text for a collection of products, instead of repeatedly using the same product information. By not allowing previous answers to be changed, and presenting translated descriptions for different products, it can help to reduce demand characteristics. This refers to the possibility that they change their responses after they realize what is being manipulated and then modify their behaviour to what they think the research demands of them (i.e. support the hypothesis). Participants often do this to please the researcher and make themselves look normal (Neuman and Neuman 2006). The countermeasure discussed makes it harder for participants to manipulate their actions to achieve their own desired outcome.

## 4　Contribution

This study is expected to contribute to the research literature on online customer shopping experience. First, the research will empirically investigate the impact of translated information quality on potential customers' satisfaction and their selection of e-retailer. A large body of prior research about e-retailing has focused on service quality, product quality, and website design, etc. However, Wolfinbarger and Gilly (2001) argue that the availability of information should be the focus of the discussion of the critical factors in customer purchase behaviour in an e-retailing environment. Considering the importance of information in an e-retailing context, the impact of information quality is worth investigating exclusively. This current research study uniquely explores the translation factor of information quality. The finding is expected to justify that the information quality is an important information system variable in an e-retailing context. This result can also further validate the information-related part of the DeLone & McLean Updated IS Success Model and it's applicability in a global e-retailing context. There are also expected to be benefits further up the supply chain for cross border B2B e-marketplaces, where the volumes of goods and amounts of money involved would place a much stronger importance upon translated information quality.

Additionally, this study contributes to theory by finding the moderating effect of brand awareness on the relation between translated information quality and perceived product risk. It helps to examine whether retailers who do not have high brand awareness, must put more effort into improving the information translation quality to alleviate the product quality risk perceived by potential customers, in order to be competitive with e-retailers which do have high brand awareness.

The research will also offer implications for practice. Few studies examine the key factors affecting foreign customers' satisfaction and intention to use a particular website, which are important when e-retailers localise their website languages for the customer's benefit. To the best of our knowledge, no previous study has examined the effect of different translation methods on customers' shopping experience. The results of this study will clarify the importance of translation. Additionally, it will suggest the level of importance for brand awareness in terms of influencing customers' tolerance of the product information quality. Based on these findings, e-retailers can evaluate whether it is worth investing to improve translation and which translation methods to choose, according to their own conditions.

## 5　References

# Copyright